\documentclass[preprint,aps]{revtex4}
\usepackage{epsfig}

\newcommand{\be}{\begin{equation}}
\newcommand{\ee}{\end{equation}}
\newcommand{\beqq}{\setlength\arraycolsep{2pt}\begin{eqnarray}}
\newcommand{\eeqq}{\vspace{0cm} \end{eqnarray}}
\newcommand{\bea}{\begin{eqnarray}}
\newcommand{\eea}{\end{eqnarray}}

\begin{document}

\title{Massless particle creation in $f(R)$ expanding universe}

\author{S. H. Pereira} \email{shpereira@gmail.com}

\affiliation{UNESP - Universidade Estadual Paulista -- Campus de Guaratinguet\'a - DFQ \\ Av. Dr. Ariberto Pereira da Cunha, 333 -- Pedregulho\\
12516-410 - Guaratinguet\'a, SP - Brazil}

\author{J. C. Z. Aguilar}

\author{E. C. Rom\~ao}

\affiliation{UNIFEI - Universidade Federal de Itajub\'a -- Campus Itabira \\ Rua Irm\~a Ivone Drumond, 200 -- CDI II \\ 35903-087 - Itabira, MG - Brazil}
   



\begin{abstract}

In this paper we present an exact solution to the spectrum of massless particle creation for a power law expansion of the scale factor of the form $t^{2n/3}$ for real scalar particles in a flat and matter dominated universe. Such an evolution follows from a modified theory of gravity of the type $f(R)=\beta R^n$, and it is showed that the spectrum of created particles is $\beta$ independent. We find that greater the value of $n$ smaller is the number of created particles. We study in detail the spectrum of the total particle number created for $n=1\pm \varepsilon$, with $\varepsilon = 0.1$, a very small deviation from the standard general relativity case ($n=1$). We find that such a very small deviation causes a great difference in the total particle number, of about one order of magnitude as compared to the standard general relativity. Our calculations are based on the method of instantaneous Hamiltonian diagonalization, where the vacuum states are defined as those which minimizes the energy at a particular instant of time. The spectrum of the total number and total energy of created particles can be determined exactly for any value of the physical time in a matter dominated universe. We also find that the main contribution to the total number of particles and total energy comes from small wavenumbers $k$ and for particles with large values of $k$ the contribution is very small. 
\end{abstract}

\maketitle

\section{Introduction}

The phenomenon of particle creation in an expanding universe has been studied by several authors \cite{davies,fulling,grib,mukh2,partcreation,staro,pavlov01,pavlov02,gribmama02,fabris01} after the pioneering work of Parker \cite{parker}. One of the most interesting results from Parker's work is that in a radiation dominated universe there is no creation of massless particles, either of zero or non-zero spin. In the description of particle production by the gravitational field, two different methods are commonly used. The standard method adopted by Parker and others \cite{davies,fulling,parker} is the method of the adiabatic vacuum state, where the vacuum state is defined as one for which the lowest-energy state goes to zero smoothly as $t\to 0$ in the past . Another widely used method is that of instantaneous Hamiltonian diagonalization \cite{grib, pavlov01} suggested by Grib and Mamayev \cite{gribmama02}, where the vacuum states are defined as those which minimizes the energy at a particular instant of time. This method has the advantage of producing a creation rate relatively higher than the previous method. Of particular interest are the works relating the gravitational particle production at the end of the inflation as a possible mechanism to produce super-heavy particles of dark matter \cite{pavlov02,superDM}. Here we restrict ourselves to this second method.

All this searches are in the framework of standard general relativity. Recently, non-standard gravity theories have been proposed as an alternative to explain the present accelerating stage of the universe only with cold dark matter, that is, with no appealing to the existence of dark energy. This is naturally obtained in $f(R)$ gravity theories (for a review, see \cite{sotiri}). In such an approach, the curvature scalar  $R$ in the Einstein-Hilbert action is replaced by a general function $f(R)$, so that the Einstein field equation is recovered as a particular case \cite{fR,allemandi,allemandi2,sotiri2,meng,dolgov,cembranos,gaspe,vilk}. A review of various modified gravities considered as gravitational alternative for dark energy, an unified description of the early-time inflation and the solutions of some related problems in $f(R)$ theories have been presented by Nojiri and Odintsov \cite{nojiri}. In a recent work \cite{saulocarlos} we have studied the quantum process of particle creation in a radiation dominated universe in view of a $f(R)=R+\beta R^n$ theory, which has the correct limit of the standard general relativity for $\beta=0$. In that paper we have showed that both massive and massless scalar particles can be produced by purely expanding effects, contrary to Parker results.

Although widely used recently as possible theories to explain the late time acceleration of the universe, $f(R)$ theories has also been severely tested for different limits, including solar systems tests \cite{solar} up to cosmic expansion history \cite{amendola}, which provides powerful constraint on
such models. For the particular $f(R)=\beta R^n$ model, it was showed in \cite{amendola} that such models can produce an expansion as $a \sim t^{2/3}$ but this does not connect to a late-time acceleration. Hence, though acceptable, the $R^n$ models are not cosmologically viable to explain the present acceleration phase of the universe.

In the present work we also study the $f(R) = \beta R^n$ in order to investigate the particle production of such model in a flat and matter dominated universe. Although not explain the current acceleration of the universe, this model is exactly solvable for the problem of particle creation. Furthermore this model provides a generalization to a scale factor of the type $a \sim t^{2n/3}$, which reproduces any phase of a FRW expansion. Quantitative calculations with $n = 9/10$ are presented, exactly the same model adopted in \cite{amendola} to discuss the viability of the model. In order to compare with the standard general relativity solution $(n=1)$, we also present the study of the model with $n=11/10$. With these choices, the scaling factor is of the form $t^{{2\over 3}(1\pm \varepsilon)}$, with $\varepsilon = 0.1$, representing a small deviation from the matter dominated universe in the standard general relativity model. Although seems to be artificial, this model allows an exact solution for the spectrum of created particles in terms of the real physical time $t$ and with initial conditions at $ t_i $.  Thus, the treatment presented here can be the basis for future research related to the creation of matter in the universe for various models of evolution. 

\section{The action of the model}

We are interested in the study of the action of the form
\be
S = S^{grav} + S^{mat}\,,
\ee
where 
\be
S^{grav}=-{1\over 16 \pi G}\int d^4x \sqrt{-g} f(R)
\ee
stands for the modified gravitational field, with $f(R)$ a function of the Ricci curvature scalar $R$, and
\begin{equation}\label{m63}
S^{mat}=\int d^4 x\sqrt{-g} \mathcal{L}^{cl} + \int d^4 x \sqrt{-g}\mathcal{L}^s\,,
\end{equation}
is the matter action, which is the sum of a classical part (composed for the barionic matter for example), represented by a Lagrangian density $\mathcal{L}^{cl}$, plus to a real massive scalar field $\phi$ minimally coupled to the gravity, represented by a Lagrangian density $\mathcal{L}^s$,
\be
\mathcal{L}^s = {1\over 2}\bigg[g^{\alpha\beta}\partial_\alpha\phi\partial_\beta\phi-m^2 \phi^2\bigg]\,.\label{lagrang}
\ee

The variation of the action $S$ with respect to $\phi$ will lead to the equation of motion for the real scalar field, namely the Klein-Gordon equation
\be
\partial_\mu\partial^{\mu}\phi +m^2 \phi = 0\,.
\ee

For the gravitational part, we will restrict ourselves to a simple $f(R)$ theory discussed in detail by Allemandi et al \cite{allemandi2} and \cite{amendola}, namely
\be\label{eq14}
f(R)=\beta R^n\,.
\ee
Taking into account the dominant energy condition, $R^n$ must be positive definite for even integer values of $n$ and $\beta$ should be fixed as $\beta >0 $ for $n<2$ and $\beta < 0$ for $n>2$. The dimension of $\beta$ is the same as the dimension of $R^{1-n}$. For $n=1$ and $\beta = 1$ we recover the standard general relativity equations. In the first order Palatine formalism, the variation of the action with respect to the Christoffel symbol $\Gamma^{\mu}_{\alpha \beta}$ (assumed independent of $g_{\mu \nu}$) gives \cite{fR}
\be
\beta R^{n-1}\Bigg[n R_{\mu \nu}-{1\over 2} R g_{\mu \nu}\Bigg]=8\pi G T_{\mu \nu}\,,\label{genR}
\ee
where $T_{\mu \nu}$ is the energy-momentum tensor of the matter fields, defined as
\be
T_{\mu\nu}\equiv {2\over \sqrt{-g}}{\delta S^{mat}\over \delta g^{\mu\nu}}= T_{\mu\nu}^{cl}+T_{\mu\nu}^s\,,
\ee
We assume that the classical part of the energy-momentum tensor can be approximated by a perfect fluid characterized by energy density $\rho^{cl}$, pressure $p^{cl}$ and four-velocity $u^{cl}$, and
\be
T_{\mu\nu}^{cl}=(\rho^{cl}+p^{cl})u_\mu u_\nu-p^{cl}g_{\mu\nu}\,.
\ee
The scalar field part of the energy-momentum tensor is
\be
T^s_{\mu\nu} = \partial_\mu \phi \partial_{\nu}\phi-{1\over 2}g_{\mu\nu}\big[g^{\alpha\beta}\partial_\alpha \phi \partial_{\beta}\phi-m^2\phi^2\big]\,.
\ee

The generalized Friedmann equations are obtained by the solution of the equation (\ref{genR}). The left hand side is a slightly modified version of the Einstein equation, which can be obtained with $\beta=n=1$. The right hand side is the energy-momentum tensor for the matter content of the universe. Note that in the case of an arbitrary scalar field the general solution is very complicated if we have no specific form for the scalar field.

\section{Particle creation in $f(R)$ theory}

The study of the quantization of a scalar field in the Hamiltonian diagonalization procedure is presented in various references \cite{grib,mukh2,staro,pavlov01,pavlov02,gribmama02}. We develop it briefly in the Appendix. The main result concerns the spectrum of the total particle number $N_k$ in a mode $k$ created as a function of the conformal time $\eta(t) \equiv \int^t dt/a(t)$, namely
\begin{equation}\label{Nk}
N_k(\eta)={1\over 4\omega_k(\eta)}|\chi'_{k}(\eta)|^2+{\omega_k(\eta)\over 4}|\chi_{k}(\eta)|^2 -{1\over 2}\,,
\end{equation}
where
\begin{equation}\label{m68}
\omega_k^2(\eta)\equiv k^2+m^2a^2(\eta)-{a''(\eta)\over a(\eta)}\,,
\end{equation}
is the squared oscillation frequency for a mode $k$ and $a(\eta)$ is the cosmological scale factor in terms of the conformal time $\eta$. The prime denotes derivatives with respect to $\eta$. The function $\chi_k(\eta)\equiv a(\eta)\phi(x)$ is the solution to the decoupled ordinary differential equations
\begin{equation}\label{nv4}
\chi''_k(\eta)+\omega_k^2(\eta)\chi_k(\eta)=0\,,
\end{equation}
with specified initial conditions at $\eta_i$.

Another important quantity is the total energy per mode, $E_k(\eta)$. It is proportional to the product of the angular frequency $\omega_k$ with the total number of created particle in that mode,
\be
E_k(\eta)=2\omega_k(\eta)N_k(\eta)\,\label{Ek}
\ee
where the factor 2 stands for the fact that the energy of particles and antiparticles are equal in this case.

Thus, the total number density of created particles $n$ and the total energy density $\varepsilon$ are readily obtained by integrating over all the modes \cite{grib,staro}
\be\label{n}
n(\eta)={1\over 2\pi^2a(\eta)^3 }\int k^2 N_k(\eta) dk\,,
\ee
\be\label{rho}
\rho(\eta)={1\over 2\pi^2a(\eta)^4 }\int k^2 E_k(\eta) dk\,.
\ee

All the expressions (\ref{Nk})-(\ref{rho}) can be rewritten in terms of the physical time $t$ by using the relation (\ref{conf}).

Before we start the calculation of the particle creation in a specific model, it is important to stress that the development presented here (and detailed in Appendix) was based on the fact that a minimum state of energy there must exist. Such minimum corresponds to $\omega_k = 0$, so that the model is valid only for $\omega_k^2 > 0$ in (\ref{m68}), since for modes of the scalar field for which the squared frequency is negative, do not oscillate and than the interpretation of the corresponding states in Hilbert space in terms of physical particles is problematic. Thus we must have $k>k_{min}\equiv a''/a - m^2a^2$ in order to have a well-defined state of lowest-energy \cite{mukh2}.

Now we will apply the above results to the study of real massless scalar particle creation for the $f(R)=\beta R^n$ theory.

In order to obtain an exact solution to the equation (\ref{genR}) for the scale factor $a(t)$, we will consider that the classical part of the energy-momentum tensor formed by dust matter, thus we have $p^{cl}=0$. For the scalar field we will assume $T^s_{\mu\nu}=0$, since in the early times no particle of the scalar fields are present, remembering that we are interested in the study of the creation of such particles.  With these assumptions, the modified Friedmann equations can be obtained \cite{allemandi2} (apart from integration constants)
\be\label{at}
a(t)= a_0 \bigg({t\over t_0}\bigg)^{2n/3}\,\,; \hspace{1cm} a_0 = \bigg({3\epsilon \over 2n(3-n)} \bigg)^{n/3} \bigg[{8\pi G \rho_0 \over \beta(2-n)} \bigg]^{1/3}.
\ee
where $\epsilon = 1$ for odd values of $n$ and $\epsilon \pm 1$ for even values.$\rho_0^{cl}$ is the present day value of the energy density of the barionic matter. Note that the case $n=3$ is singular.

To follow up we need the scale factor $a(t)$ in terms of the conformal time $\eta$. This can be obtained by integrating (\ref{conf}) and rewritten $a(t)$ in terms of $\eta$,
\be 
a(\eta)= a_1 \eta^{2n\over 3-2n}\,\,; \hspace{1cm} a_1=a_0\Big[a_0\bigg({3-2n\over 3}\bigg)^{2n\over 3-2n} \Big] \,\,; \hspace{1cm} 0<\eta<\infty \,.\label{aeta}
\ee
with $0<\eta<\infty$ for $0<n<3/2$ and $-\infty<\eta<0$ making $\eta \to -\eta$ in (\ref{aeta}) for $n>3/2$ or $n<0$. This last condition implies that in terms of the conformal time, the early universe (the past) corresponds to $\eta\to -\infty$ and the late universe (the future) corresponds to $\eta \to 0$.

For the massless case ($m=0$), substituting (\ref{aeta}) in the frequency mode (\ref{m68}) gives
\be
\omega_k(\eta) = \sqrt{k^2-{p(p-1)\over \eta^2}}\,\,;\hspace{1cm} p={2n\over 3-2n}\,,\label{omp}
\ee
which shows that even in the massless case, the scalar field is endowed with an negative effective mass. A negative value for the effective mass can arise because of the field interaction with the gravitational background \cite{mukh2}.

The equation (\ref{nv4}) for the mode function is
\be
\chi_k''(\eta)+\bigg[k^2-{p(p-1)\over \eta^2}\bigg]\chi_k(\eta)=0\,\,.\label{modeeq}
\ee
A very interesting characteristic of this equation is that it does not depends on the $\beta$ parameter, which shows that its solution is general and the spectrum of created particles does not change for an arbitrary value of $\beta$.

In order to satisfy the initial conditions (\ref{inicond}), the general solution of the above equation is given by
\be\label{solution}
\chi_{k}(\eta)=A(k,\eta_i)\sqrt{\eta}J_{p-{1\over 2}}(k\eta) - B(k,\eta_i)\sqrt{\eta}Y_{p -{1\over 2}}(k\eta)\,,
\ee
where $J_\nu$ and $Y_\nu$ are the Bessel functions of first and second kind, respectively, $\nu$ is its order and $A$ and $B$ are complex constants depending on $k$ and on the initial time $\eta_i$,
\beqq
A(k,\eta_i)={k\eta_iY_{p+{1\over2}}(k\eta_i)-p Y_{p-{1\over2}}(k\eta_i)+ I\eta_i \omega_k(\eta_i) Y_{p-{1\over2}}(k\eta_i)\over k\eta_i^{3/2}\sqrt{\omega_k(\eta_i)}[J_{p-{1\over2}}(k\eta_i)Y_{p+{1\over2}}(k\eta_i)-J_{p+{1\over2}}(k\eta_i)Y_{p-{1\over2}}(k\eta_i) ] }\,,\nonumber\\
B(k,\eta_i)={k\eta_iJ_{p+{1\over2}}(k\eta_i)-p J_{p-{1\over2}}(k\eta_i)+ I\eta_i \omega_k(\eta_i) J_{p-{1\over2}}(k\eta_i)\over k\eta_i^{3/2}\sqrt{\omega_k(\eta_i)}[J_{p-{1\over2}}(k\eta_i)Y_{p+{1\over2}}(k\eta_i)-J_{p+{1\over2}}(k\eta_i)Y_{p-{1\over2}}(k\eta_i) ] }\,,
\eeqq
where $\omega_k(\eta_i)=\sqrt{k^2-p(p-1)/\eta_i^2}$ and $I\equiv \sqrt{-1}$.

By fixing $\eta_i$ and taking some particular values of $k$ and $\eta$ it is easy to check that this solution correctly satisfy the normalization condition (\ref{norm}).

The solution presented in (\ref{solution}) is exact and, at first, it may be used to calculate the total number density of particles (\ref{n}) and the total energy density (\ref{rho}) for any instant of time $\eta$ after the initial time $\eta_i$. To recover the expressions as a function of physical time $t$ we use (\ref{conf}) and all quantities can be described in terms of $t$. In fact, the final analytical expression for $N_k$ is quite complicated, but can be obtained in relatively simple manner with the help of an algebraic manipulation software. Thus, the exact behavior and several interesting features of the functions (\ref{Nk}) - (\ref{rho}) can be obtained by graphical analysis of the spectrum of created particles. This is what we will do in the next section.



\section{Spectrum of created particles}

Let us study the evolution in time of the total particle number $N_k$. First we need to establish an appropriate time scale in order to make the analytical study. The present time corresponds to $t_0 = H_0^{-1} = 4.4 \times 10^{17}$s, so that an appropriate time scale is in terms of $H_0^{-1}$. Now we will establish the value of the initial time $t_i$. As we are dealing with a universe dominated by matter and we know \cite{kolb} that the matter and radiation decouple at about $t_i = 10^{13}$s $\approx 4.5 \times 10^{-5} H_0^{-1}$, we will use such value for the initial time. Thus we are interested in the interval $4.5\times 10^{-5}H_0^{-1} < t < H_0^{-1}$, which corresponds to the matter era.

Now we can replace (\ref{solution}) into (\ref{Nk}) to obtain the spectrum of the total number of particles created for each mode $k$, for a given power law expansion (\ref{at}). Since the solution of (\ref{modeeq}) does not depends on the $a_0$ factor, it does not depends on the $\beta$ parameter too, thus our result is general. For the standard general relativity we have $n=1$ and the solution (\ref{at}) corresponds to the matter dominated expansion. For the modified gravity we choose the model $n=9/10$, discussed in detail in \cite{amendola}, and the model $n=11/10$, in order to study a small deviation from the standard model, $n=1$. We use (\ref{conf}) to write the result as a function of physical time $t$. The spectrum of the total number of created particles $N_k$ for some values of the $k$ mode is represented in the Figure 1-(a). By analyzing the Figure we see that for every mode $k$ there is a abrupt growth of the particle number at the initial time and after a brief oscillation it turns constant, and remains so throughout evolution. The behavior at $t_i$ is consistent with the initial conditions (\ref {inicond}) that minimizes the energy at that point. Another feature is that the number of particles created decreases with increasing $k$. Exactly the same behavior is obtained for the total energy $E_k(t)$ given by (\ref{Ek}), since it is proportional to $N_k$. A much more interesting result is that for $n=1-\varepsilon$ we see that the total number of created particles increases as compared to the $n=1$ case, and such increasing is about one order of magnitude for each mode, thus we conclude that a small deviation from the standard general relativity model leads to a great variation in the particle number. The same behaviour is obtained for $n=1+\varepsilon$, but in this case the particle total number reduces by one order of magnitude for each mode. We also conclude that greater the value of the deviation from the standard case smaller is the total number of created particle.

\begin{figure}[tb]
\begin{center}
\epsfig{file=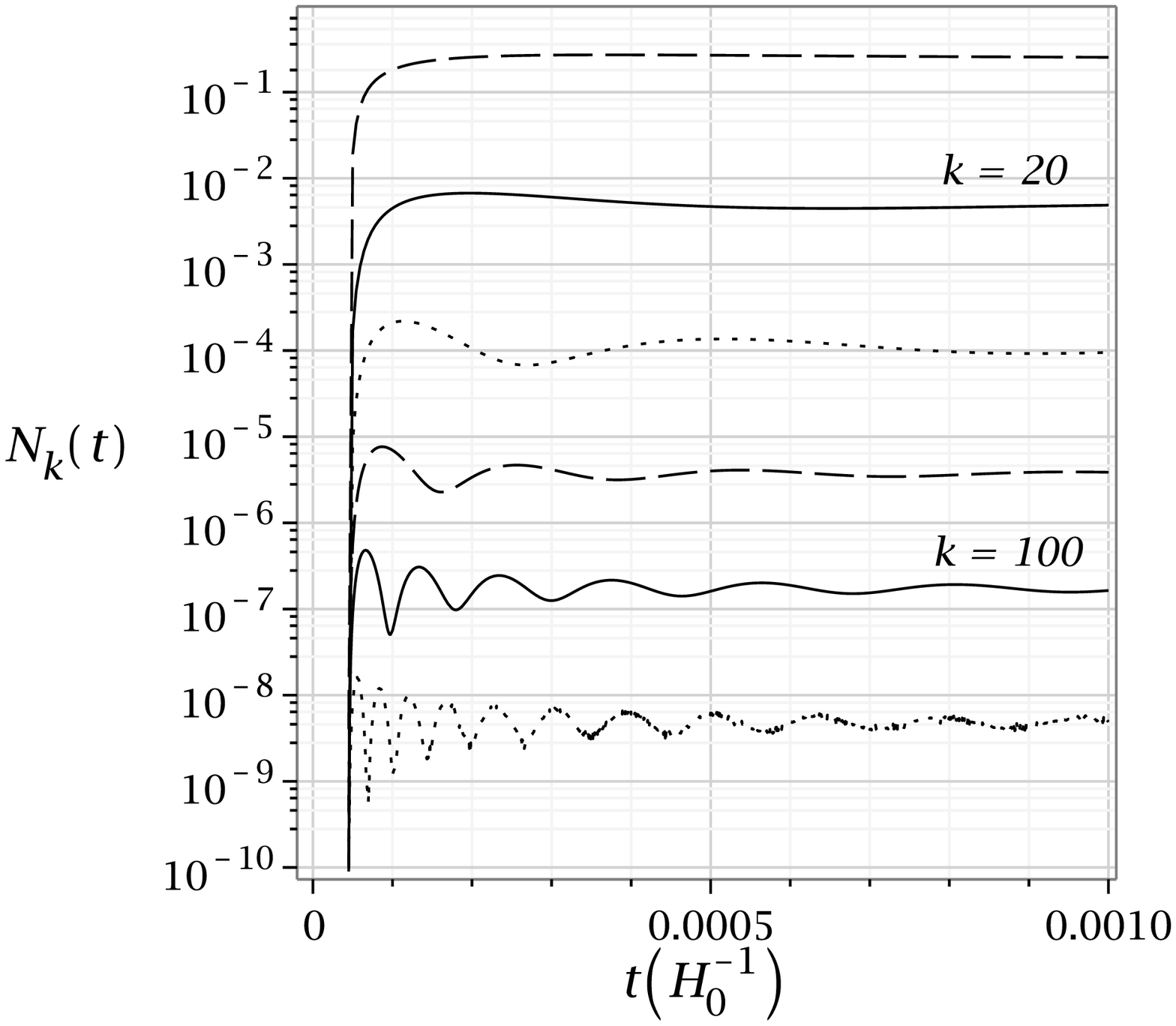, scale=0.37} \hspace{1cm}\epsfig{file=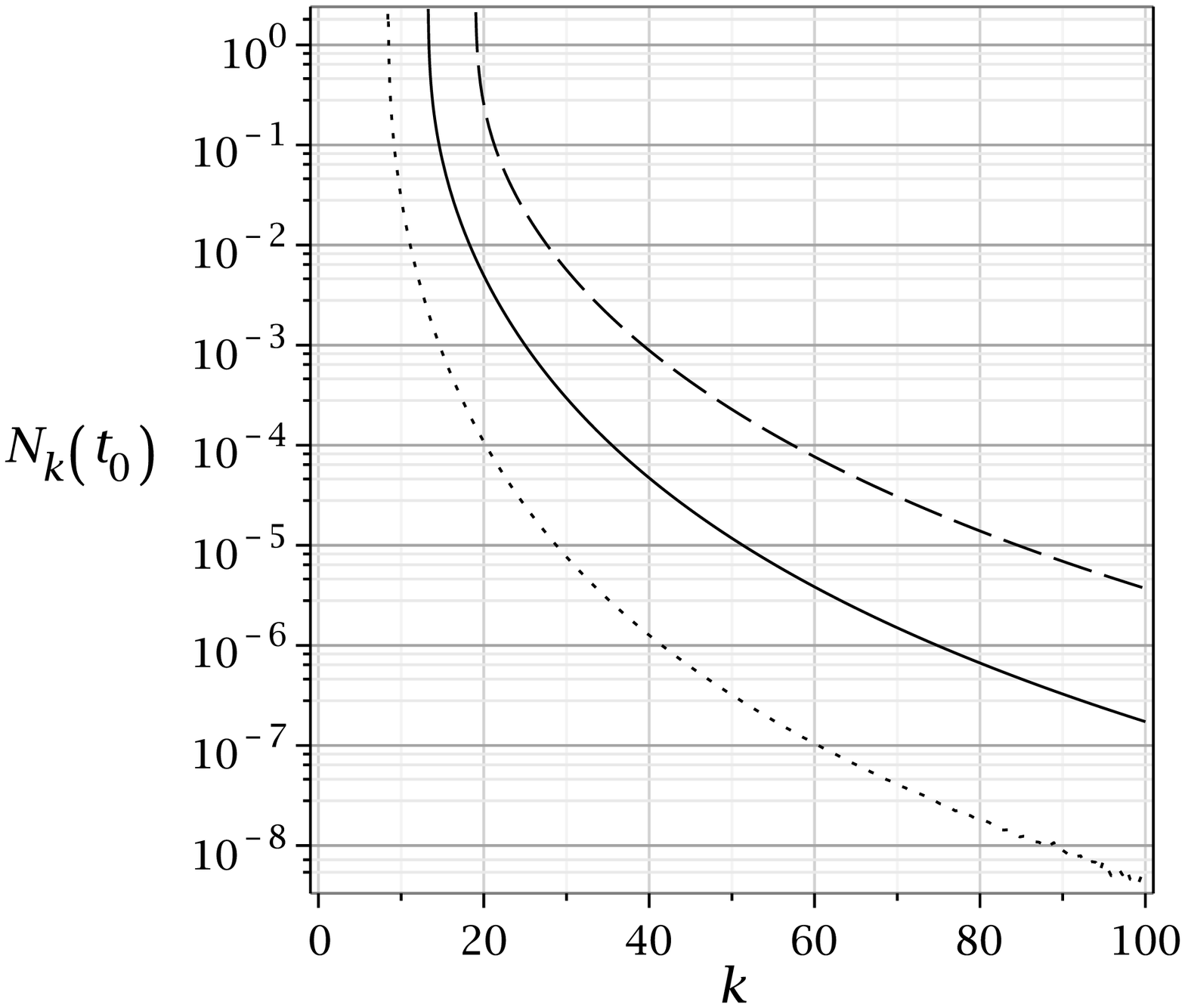, scale=0.37}\\
\hspace{1cm}(a)\hspace{8.2cm}(b)
\end{center}
\caption{(a) - Spectrum of the total particle number $N_k$ as a function of the physical time for the standard cosmology (solid line), for $f(R)=\beta R^{0.9}$ (dashed) and $f(R)=\beta R^{1.1}$ (dotted), for the modes $k=20$ (upper three) and $k=100$ (lower three); (b) - the total particle number for the present time $t_0$ as a function of the $k$ mode, for the standard cosmology (solid line), for $f(R)=\beta R^{0.9}$ (dashed) and $f(R)=\beta R^{1.1}$ (dotted).}\label{fig1}
\end{figure}

Another interesting analysis that we can do based on the analytical results is the behavior of $N_k$ as a function of $k$ for some fixed time. This is shown in Figure 1-(b) for the present time $t = t_0$, for the standard model ($n=1$, solid line) and the two models of $f(R)$ gravity (dashed and dotted lines). The total number of particles per mode appears to diverges for a given initial value of the mode $k$ for all models, but decreases with increasing $k$. The decrease of the total particle number with the increase of the mode $k$ is an interesting feature, indicating that higher modes contribute much less to the total number of particles and the total energy. By fixing a mode $k$ we also verify that the total number increases by about 10 times from $n+\varepsilon \to n$ and from $n \to n-\varepsilon$.

As we have already discussed before, the model is valid only for modes $k > k_{min}$, and $k_{min}$ must be taken as that corresponding to the minimal value of the time $t_i$ when inserted in the $\omega_k$ expression (\ref{m68}). This explain the divergent value for the modes in the Figure 1-(b). For the standard cosmology ($n=1$) such value is $k_{min}=13.25$, for $n=9/10$ it is $k_{min}=18.98$ and for $n=11/10$ it is $k_{min}=8.44$. These behaviour is because only the modes with $k > k_{min}$ are present in the field, since the modes with $k<k_{min}$ does not have a  well-defined lowest-energy state. 

Mathematically the divergence in $k$ occurs because in the expression (\ref{solution}) the term $\omega_k(\eta_i)$ appears in the denominator, thus it diverges for some minimum value of $k$. Fortunately, such divergence is removed when the integration of $k$ starts on $k_{min}$, as must be in our case. Thus, the integral of the total particle number and the total energy is finite.

\section{Concluding remarks}

We have addressed the problem of minimally coupled real massless scalar particle creation in a $f(R)=\beta R^n$ theory for a flat and matter dominated universe. Such specific form of $f(R)$ theory leads to a scale factor evolution of the form $a\sim t^{2n/3}$. We present an exact solution for the spectrum of total particle number with initial conditions that minimizes the energy, based on the instantaneous Hamiltonian diagonalization method. For the specific case  $n=1 + \varepsilon$ we have found that a small deviation $\varepsilon$ from the standard general relativity model ($n=1$) leads to a great increasing or decreasing in the total particle number if $\varepsilon = -0.1$ or $\varepsilon = +0.1$, respectively. Such a difference is of about one order of magnitude. We also have found that the main contribution to the spectrum of created particles comes from small values of the mode $k$.  The study presented here can be the basis for future researches related to the creation of matter in the universe for various models of evolution.
  
\appendix
\section*{Appendix}

The canonical quantization of a real minimally coupled scalar field in  curved backgrounds follows in straight analogy with the quantization in a flat Minkowski background. The gravitational metric is treated as a classical external field which is generally non-homogeneous and non-stationary. The action for a massive real scalar field minimally coupled to gravity in a spatially flat Friedmann-Robertson-Walker geometry is \cite{mukh2}:
\be
S^s= {1\over 2}\int d^4 x \sqrt{-g}\bigg[g^{\alpha\beta}\partial_\alpha\phi\partial_\beta\phi-m^2 \phi^2\bigg]\,.\label{action1}
\ee
With the substitution $\phi({\bf x},t)=\chi({\bf x},\eta)a(\eta)$, $g^{\alpha\beta}=a^{-2}\eta^{\alpha\beta}$ and $\sqrt{-g}=a^4$, where $\eta^{\alpha\beta}$ is the flat Minkowski metric, and taking the conformal time $\eta$ and the physical time $t$ related by
\be
\eta \equiv \int {dt\over a(t)}\,,\label{conf}
\ee
the action (\ref{action1}) can be rewrited as
\be
S^s= {1\over 2}\int d^3 {\bf x} \,d\eta\bigg[\chi'^2 - \big(\nabla \chi \big)^2-\Big(m^2a^2-{a''\over a} \Big)\chi^2\bigg]\,,\label{action2}
\ee
where $a(\eta)$ is the cosmological scale factor in terms of the conformal time $\eta$ and the prime denotes derivatives with respect to it. The variation of this action with respect to $\chi$ gives the equation of motion for the field $\chi$: 
\be
\chi'' - \nabla^2\chi +\Big(m^2a^2-{a''\over a} \Big)\chi =0\,.
\ee
Following standard lines, the quantization of the scalar field can be carried out by promoting $\chi$ to a operator $\hat{\chi}$ and imposing equal-time commutation relations for the scalar field operator $\hat{\chi}$ and its canonically conjugate momentum $\hat{\pi}\equiv\hat{\chi}'$, namely $[\hat{\chi}({\bf x},\eta) \,, \hat{\pi}({\bf y},\eta)]=i\delta({\bf x}-{\bf y})$ (and zero for the others), and by implementing secondary quantization. 

Expanding the field operator $\hat{\chi}$ in Fourier modes, 
\be
\hat{\chi}({\bf x},\eta)={1\over \sqrt{2}}\int{d^3{\bf k}\over (2\pi)^{3/2}}\bigg[b_k^-\chi_k^*(\eta)\exp(i{\bf k}\cdot{\bf x}) + b_k^+\chi_k(\eta)\exp(-i{\bf k}\cdot{\bf x})\bigg]\,,
\ee
where $b_k^+$ and $b_k^-$ are the creation and annihilation operators of the field, respectively. The equation of motion for the decoupled modes $\chi_k(\eta)$ is
\begin{equation}\label{nv4}
\chi''_k(\eta)+\omega_k^2(\eta)\chi_k(\eta)=0\,,
\end{equation}
with
\begin{equation}\label{m68}
\omega_k^2(\eta)\equiv k^2+m^2_{eff} \hspace{1cm} \textrm{and} \hspace{1cm} m^2_{eff}\equiv m^2a^2(\eta)-{a''(\eta)\over a(\eta)}\,,
\end{equation}
where $k$ is the Fourier mode or wavenumber of the particle, $\omega_k$ and $m_{eff}$ represents the frequency and the effective mass of the particle, respectively.  The operators $b_k^\pm$ satisfy the commutation relations $[b_k^-\,, b_{k'}^+]=\delta({\bf k}-{\bf k}')$ (and zero for the others).
 
To proceed, note that Eq. (\ref{nv4}) is a second order differential equation with two independent solutions. Each solution $\chi_{k}$ must be normalized for all times according to
\begin{equation}\label{norm}
W_k(\eta)\equiv \chi_{k}(\eta){\chi^*}'_{k}(\eta)-\chi'_{k}(\eta)\chi^*_{k}(\eta)=-2i\,,
\end{equation}

The Hamiltonian operator can be writing as
\be
\hat{H}(\eta)={1\over 4}\int d^3{\bf k}\bigg[b_k^-b_{-k}^- F^*_k + b_k^+b_{-k}^+ F_k+ \big(b_k^+b_k^-+b_k^-b_k^+\big)E_k\bigg]\,,\label{ham}
\ee
where
\be
E_k(\eta)=|\chi_k'|^2+\omega_k^2(\eta)|\chi_k|^2\,, \hspace{1cm}F_k(\eta)=\chi_k'^2+\omega_k^2(\eta)\chi_k^2\,.
\ee
As can be seen, the Hamiltonian is non-diagonal in the creation and annihilation operators, which leads to the creation of particle-antiparticle pairs from the vacuum. If we assume that the scalar field is in the vacuum state at a certain initial time $\eta_i$, we must have $E_k(\eta_i)=1$ and $F_k(\eta_i)=0$. These conditions are satisfied if we have the following initial conditions at the time $\eta_i$:
\be
\chi_k(\eta_i)=1/\sqrt{\omega_k(\eta_i)}\,,\hspace{1cm} \chi_k'(\eta_i)=i\sqrt{\omega_k(\eta_i)}\,.\label{inicond}
\ee

The Hamiltonian diagonalization for an arbitrary instant $\eta$ is realized in terms of a time-dependent Bogolyubov transformation of the operators $b_k^\pm$ to new operators $c_k^\pm(\eta)$,
\be
b_k^-=\alpha_k^*(\eta) c_{k}^-(\eta) +\beta_k(\eta)c_{-k}^+(\eta)\,\,; \hspace{1cm} b_k^+=\alpha_k^*(\eta) c_k^+(\eta) +\beta_k^*(\eta)c_{-k}^-(\eta)
\ee
with $|\alpha_k|^2-|\beta_k|^2=1$. Thus, the Hamiltonian can be writing as
\be
\hat{H}(\eta)=\int d^3{\bf k}\,\omega_k(\eta)\bigg[{1\over 2}+ |\beta_k(\eta)|^2\bigg]\,.
\ee
This result allows us to write the total number of created particles $N_k$ and antiparticles $\bar{N}_k$ as
\begin{equation}\label{Nk}
N_k(\eta)=\bar{N}_k(\eta)\equiv |\beta_k(\eta)|^2={1\over 4\omega_k(\eta)}|\chi'_{k}(\eta)|^2+{\omega_k(\eta)\over 4}|\chi_{k}(\eta)|^2 -{1\over 2}\,.
\end{equation}


\begin{acknowledgements}
SHP is grateful to CNPq - Conselho Nacional de Desenvolvimento Cient\'ifico e Tecnol\'ogico, Brazilian research agency, for the financial support, process number 477872/2010-7.
\end{acknowledgements}

\end{document}